\documentclass{article}
\usepackage{spconf,amsmath,graphicx}
\usepackage[moderate]{savetrees}
\usepackage{epigraph}
\usepackage{amssymb}
\usepackage{pifont}
\usepackage{multirow}
\usepackage{footmisc}
\usepackage{subcaption}
\usepackage{makecell}
\usepackage{listings}
\usepackage{xcolor}
\usepackage{xspace}
\usepackage{stix}
\usepackage{dirtytalk}
\usepackage{amssymb}
\usepackage{amsmath}
\usepackage{amsfonts}
\usepackage{booktabs}
\usepackage{algorithm}
\usepackage{algpseudocode}

\usepackage{hyperref}
\hypersetup{
    colorlinks=true,
    linkcolor=purple,
    filecolor=magenta,      
    urlcolor=purple,
}

\usepackage[numbers,sort]{natbib}
\newcommand{\xmark}{\ding{55}}%
\newcommand{\cmark}{\ding{51}}%

\usepackage[utf8]{inputenc}
\usepackage[title]{appendix}

\definecolor{codegreen}{rgb}{0,0.6,0}
\definecolor{codegray}{rgb}{0.5,0.5,0.5}
\definecolor{codepurple}{rgb}{0.58,0,0.82}
\definecolor{backcolour}{rgb}{0.95,0.95,0.92}
 
\lstdefinestyle{mystyle}{
    backgroundcolor=\color{backcolour},   
    commentstyle=\color{codegreen},
    keywordstyle=\color{magenta},
    numberstyle=\tiny\color{codegray},
    stringstyle=\color{codepurple},
    basicstyle=\ttfamily\footnotesize,
    breakatwhitespace=false,         
    breaklines=true,                 
    captionpos=b,                    
    keepspaces=true,                 
    numbers=left,                    
    numbersep=5pt,                  
    showspaces=false,                
    showstringspaces=false,
    showtabs=false,                  
    tabsize=2,
    basicstyle=\fontsize{7}{8}\ttfamily
}
 
\title{OxfordVGG Submission to the EGO4D AV Transcription challenge}
\name{Jaesung Huh, Max Bain, Andrew Zisserman}
\address{Visual Geometry Group, University of Oxford, UK}

\begin{document}
%
\maketitle
\begin{abstract}
This report presents the technical details of our submission on the EGO4D Audio-Visual (AV) Automatic Speech Recognition Challenge 2023 from the OxfordVGG team. 
We present \textit{WhisperX}, a system for efficient speech transcription of long-form audio with word-level time alignment, along with two text normalisers which are publicly available.
Our final submission obtained 56.0\% of the Word Error Rate (WER) on
the challenge test set, ranked 1st on the leaderboard.
All baseline codes and models are available on \url{https://github.com/m-bain/whisperX}.

\end{abstract}

\section{Introduction}
Speech recognition has been a fundamental challenge in the field of audio processing, aiming to convert speech waveforms into textual representations. 
In recent years, Deep Neural Networks (DNNs) have significantly advanced the field by improving the performance of speech recognition systems. 
These improvements have been achieved either through the hybrid DNN-HMM architecture~\cite{morgan1990continuous, mohamed2011acoustic} or by leveraging end-to-end models~\cite{Chan15, gulati2020conformer, baevski2020wav2vec}.

The availability of web-scale datasets and advancements in semi- or unsupervised learning techniques have further propelled the performance of speech recognisers to new heights. 
Notably, Whisper~\cite{radford2022robust} shows that even a simple encoder-decoder architecture could generalise well training with 680,000 hours of data.
However, it is worth noting that Whisper's input window is limited to audio segments of only 30 seconds. 
As a result, it faces challenges when transcribing longer audio files. 
Additionally, due to its sequential decoding approach, Whisper is susceptible to issues such as hallucinations or repetitive outputs.
These challenges are similar to those encountered in auto-regressive language generation tasks.

\textit{WhisperX}~\cite{bain2022whisperx} proposes a method to improve both the accuracy and efficiency of Whisper when transcribing long audio.
It uses a voice activity detection model to pre-segment the input audio to run Whisper with a cut\&merge scheme, allowing long-form audio to be transcribed in parallel with batched transcription of the pre-segmented audio chunks.
It also conducts forced phoneme alignment using an off-the-shelf model such as Wav2Vec2~\cite{baevski2020wav2vec} to generate word-level timestamps required by the EGO4D transcription challenge.

This report investigates how \textit{WhisperX} performs on the EGO4D speech transcription dataset.
EGO4D challenge dataset presents unique difficulties on two fronts. 
Firstly, unlike other widely used speech datasets~\cite{panayotov2015librispeech, bu2017aishell}, it comprises audio recordings captured in real-world scenarios with diverse types of background noise. 
Secondly, the audio files in the EGO4D dataset are recorded using a microphone positioned on the wearer's head-mounted camera. 
The frequent movements of the wearer introduce variations in the audio amplitude, making the transcription process more difficult.
Our model achieved a Word Error Rate (WER) of 56.0\% on the challenge test set, showing a substantial improvement over the baseline results.
Additionally, we show the significance of text normalisation in achieving favourable WER outcomes.

\section{Method}
\label{sec:method}
Here, we present our model \textit{WhisperX} and the text normalisation methods we've used for submission. 
Please refer to the original paper for more details~\cite{bain2022whisperx}.

\subsection{WhisperX}
\label{subsec:whisperx}

WhisperX is a time-accurate speech recognition system enabling within-audio parallelised transcription.
It employs several pre-processing steps.

The input audio is first segmented with Voice Activity Detection (VAD) model, divided into a set of voice regions.
These regions are then cut \& merged into approximately 30-second input chunks.
The paper shows that this VAD Cut \& Merge preprocessing reduces hallucination and repetition and enables within-audio batched transcription.
The resulting chunks are then: (i) transcribed in parallel with Whisper, and (ii) aligned with a phoneme recognition model to generate precise timestamps for each word. 

In our submission, we use \texttt{pyannote}~\cite{Bredin2020} VAD model and Wav2Vec2~\cite{baevski2020wav2vec} fine-tuned with 960 hours of Librispeech~\cite{panayotov2015librispeech} for phoneme alignment.
Unless specified, we follow the default settings outlined in the original paper~\cite{bain2022whisperx}.

\begin{figure*}[t]
    \centering
    \vspace{-20pt}
    \includegraphics[width=0.95\textwidth]{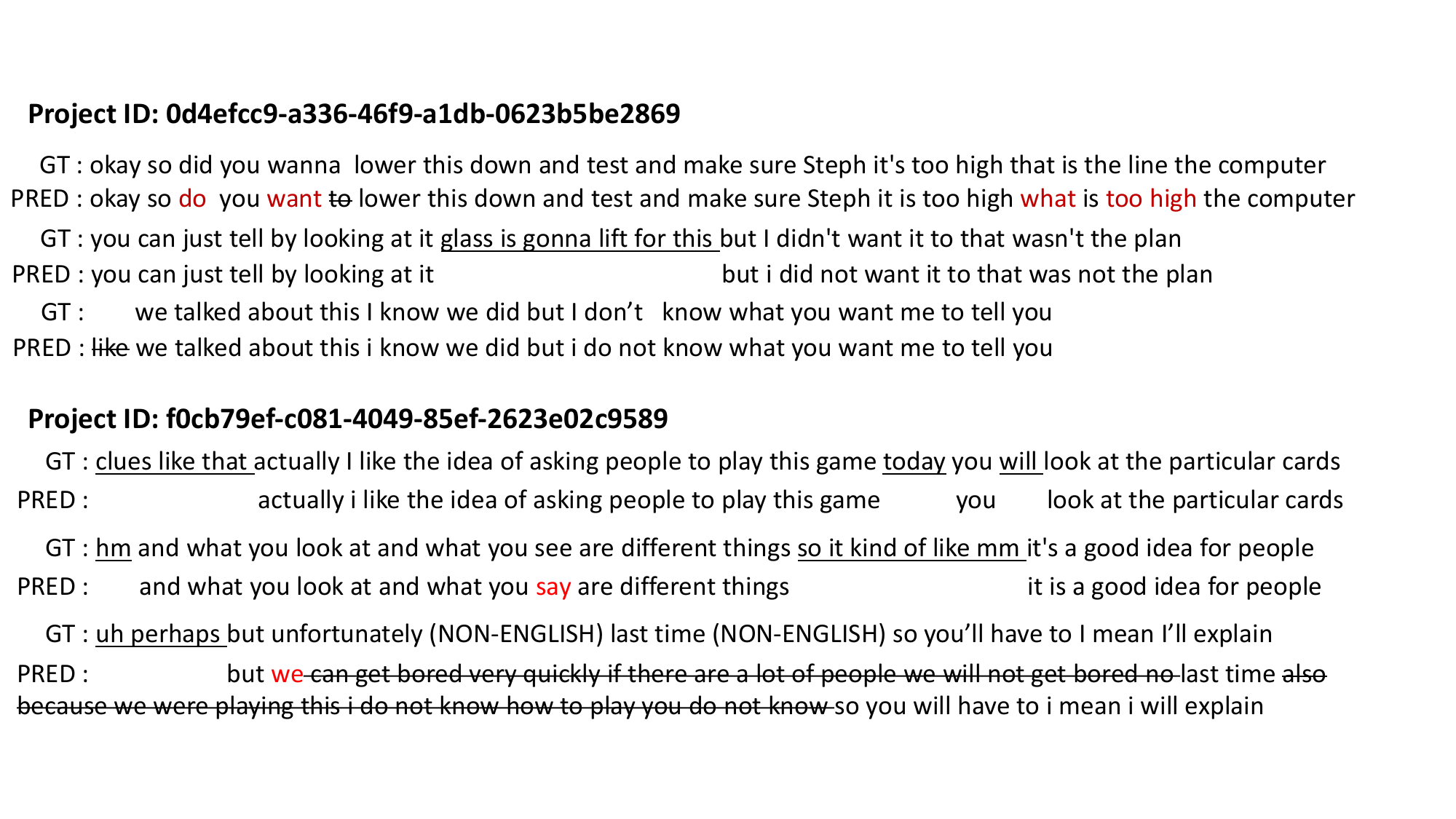}
    \vspace{-20pt}
    \caption{Qualitative examples. The figure shows the first few lines of results from audio file which produces the best WER (top) and the worst WER (bottom). \textbf{GT}: ground truth, \textbf{PRED} : our model's prediction.}
    \label{fig:qual}
\end{figure*}

\subsection{Text normaliser}
\label{subsec:text_normalisation}
Although the challenge evaluation script normalises the submission results using the English global mapping file from the Kaldi repository~\cite{kaldiglm}, additional post-processing steps are essential to improve our performance.
For example, WhisperX normally outputs the number as a form of integer or float, but the ground truth in the validation set transcribes the number into spoken forms. (e.g. 1 to `one'). 
For this reason, we use two text normalisers, Whisper text normaliser~\cite{radford2022robust} and NeMo text normaliser~\cite{zhang21ja_interspeech}.

After obtaining the initial transcript from \textit{WhisperX}, we utilise the text normaliser provided in Whisper's original repository~\cite{whisper_repo}, except keeping the interjections such as 'hmm' or 'oh'.
These interjections are not ignored in the validation set, whereas the original Whisper normaliser does overlook them. 
We also observe that Whisper normaliser converts numbers from words to digits, whereas the numbers in validation set are represented as words. 
Therefore, we employ the NeMo text normaliser to process the output of the Whisper normaliser, converting the numbers into their corresponding word forms.
This significantly improves the WER in both the validation and the test set, shown in Table~\ref{tbl:val_result} and ~\ref{tbl:test_result}.

\section{Result}
Table~\ref{tbl:val_result} shows the result on the validation set with different versions of Whisper and WhisperX.
We can see that using the text normaliser results in a significant improvement on the WER.
WhisperX also performs better than Whisper in general due to reduced hallucinations and repetitions.

Interestingly, \texttt{medium.en} performs best on the validation set.
The \texttt{medium.en} model assumes that the language spoken in the utterance is purely English, whereas the \texttt{large-v2} model checks the first 30 seconds of the audio to determine the language spoken in the whole conversation, and assumes this when making a transcript, which may lead to non-English results.
In fact, we notice that out of 50 WhisperX outputs in the val set, 13 are transcribed with non-English.
However, all the audio files in the test set are detected as English, so we decide to use \texttt{large-v2} for its greater capacity and generalisation.

Table~\ref{tbl:test_result} shows the result on the challenge test set. 
We could see using normaliser results in significant improvement in WER (73.3 to 56.0).

\begin{table}[!ht]
\centering
\resizebox{0.9\linewidth}{!}{
\begin{tabular}{llcc}
\toprule
Model & Version & Before normaliser & After normaliser\\ \midrule
\multirow{3}{*}{Whisper}& \texttt{base.en} & 74.4 & 68.2 \\
& \texttt{medium.en} & 72.0 & 65.6 \\
& \texttt{large-v2} & 75.3  & 73.3  \\ \midrule
\multirow{3}{*}{WhisperX}& \texttt{base.en} & 73.8 & 71.0 \\
& \texttt{medium.en} & 67.0 & 62.0 \\
& \texttt{large-v2} & 73.7  & 73.3  \\ \bottomrule
\end{tabular}}
\caption{Word error rate (\%) on validation set. The lower is better.}
\label{tbl:val_result}
\end{table}
\begin{table}[!ht]
\centering
\begin{tabular}{lcc}
\toprule
Version & Normaliser & WER \\ \midrule
\multirow{2}{*}{\texttt{large-v2}} & \xmark & 73.3  \\ 
 & \cmark & \textbf{56.0}  \\ \bottomrule
\end{tabular}
\caption{Word error rate (\%) on test set. The lower is better.}
\label{tbl:test_result}
\end{table}

\subsection{Qualitative Examples}
\label{subsec:qual}
Figure~\ref{fig:qual} shows some of the qualitative examples of our method along with the ground truth.
We show the project with the best WER (0d4efcc9-a336-46f9-a1db-0623b5be2869) and the worst WER (f0cb79ef-c081-4049-85ef-2623e02c9589) in the validation set.
In the worst example, the speakers speak a non-English language in the middle of the conversation (\texttt{NON-ENGLISH} in Figure~\ref{fig:qual}), which is not transcribed in the ground truth.
However, our model assumes that English is being spoken, resulting in a hallucination that causes a huge insertion error.

\section{Conclusion and Future work}
\label{sec:conclusion}
Here we report our results using WhisperX and text normalisation in the EGO4D AV transcription challenge. 
All the code and models we've used are publicly available at \url{https://github.com/m-bain/whisperX}.
Note that our method does not use visual streams, which has been shown to be helpful in recent works~\cite{Afouras20, ma2023auto}.
Also, as shown in our qualitative example, automatic language detection + multilingual speech recognition could be helpful to improve speech recognition performance.

\bibliographystyle{IEEEtranS}
\bibliography{shortstrings,vgg_local,mybib,vgg_other}

\begin{thebibliography}{10}
\providecommand{\url}[1]{#1}
\csname url@samestyle\endcsname
\providecommand{\newblock}{\relax}
\providecommand{\bibinfo}[2]{#2}
\providecommand{\BIBentrySTDinterwordspacing}{\spaceskip=0pt\relax}
\providecommand{\BIBentryALTinterwordstretchfactor}{4}
\providecommand{\BIBentryALTinterwordspacing}{\spaceskip=\fontdimen2\font plus
\BIBentryALTinterwordstretchfactor\fontdimen3\font minus
  \fontdimen4\font\relax}
\providecommand{\BIBforeignlanguage}[2]{{%
\expandafter\ifx\csname l@#1\endcsname\relax
\typeout{** WARNING: IEEEtranS.bst: No hyphenation pattern has been}%
\typeout{** loaded for the language `#1'. Using the pattern for}%
\typeout{** the default language instead.}%
\else
\language=\csname l@#1\endcsname
\fi
#2}}
\providecommand{\BIBdecl}{\relax}
\BIBdecl

\bibitem{kaldiglm}
``Kaldi english glm file,''
  \url{https://github.com/kaldi-asr/kaldi/blob/master/egs/ami/s5/local/english.glm},
  2022.

\bibitem{Afouras20}
T.~Afouras, J.~S. Chung, and A.~Zisserman, ``Asr is all you need: Cross-modal
  distillation for lip reading,'' in \emph{International Conference on
  Acoustics, Speech, and Signal Processing}, 2020.

\bibitem{baevski2020wav2vec}
A.~Baevski, Y.~Zhou, A.~Mohamed, and M.~Auli, ``wav2vec 2.0: A framework for
  self-supervised learning of speech representations,'' \emph{Advances in
  neural information processing systems}, vol.~33, pp. 12\,449--12\,460, 2020.

\bibitem{bain2022whisperx}
M.~Bain, J.~Huh, T.~Han, and A.~Zisserman, ``Whisperx: Time-accurate speech
  transcription of long-form audio,'' \emph{Proc. Interspeech 2023}, 2023.

\bibitem{Bredin2020}
H.~{Bredin}, R.~{Yin}, J.~M. {Coria}, G.~{Gelly}, P.~{Korshunov},
  M.~{Lavechin}, D.~{Fustes}, H.~{Titeux}, W.~{Bouaziz}, and M.-P. {Gill},
  ``{pyannote.audio: neural building blocks for speaker diarization},'' in
  \emph{ICASSP}, 2020.

\bibitem{bu2017aishell}
H.~Bu, J.~Du, X.~Na, B.~Wu, and H.~Zheng, ``Aishell-1: An open-source mandarin
  speech corpus and a speech recognition baseline,'' in \emph{2017 20th
  conference of the oriental chapter of the international coordinating
  committee on speech databases and speech I/O systems and assessment
  (O-COCOSDA)}.\hskip 1em plus 0.5em minus 0.4em\relax IEEE, 2017, pp. 1--5.

\bibitem{Chan15}
W.~Chan, N.~Jaitly, Q.~V. Le, and O.~Vinyals, ``Listen, attend and spell,''
  \emph{arXiv preprint arXiv:1508.01211}, 2015.

\bibitem{gulati2020conformer}
A.~Gulati, J.~Qin, C.-C. Chiu, N.~Parmar, Y.~Zhang, J.~Yu, W.~Han, S.~Wang,
  Z.~Zhang, Y.~Wu \emph{et~al.}, ``Conformer: Convolution-augmented transformer
  for speech recognition,'' \emph{arXiv preprint arXiv:2005.08100}, 2020.

\bibitem{whisper_repo}
J.~W. Kim, ``Whisper word-level timestamps,''
  \url{https://github.com/openai/whisper/blob/word-level-timestamps/notebooks/Multilingual_ASR.ipynb},
  2022.

\bibitem{ma2023auto}
P.~Ma, A.~Haliassos, A.~Fernandez-Lopez, H.~Chen, S.~Petridis, and M.~Pantic,
  ``Auto-avsr: Audio-visual speech recognition with automatic labels,'' in
  \emph{ICASSP 2023-2023 IEEE International Conference on Acoustics, Speech and
  Signal Processing (ICASSP)}.\hskip 1em plus 0.5em minus 0.4em\relax IEEE,
  2023, pp. 1--5.

\bibitem{mohamed2011acoustic}
A.-r. Mohamed, G.~E. Dahl, and G.~Hinton, ``Acoustic modeling using deep belief
  networks,'' \emph{IEEE transactions on audio, speech, and language
  processing}, vol.~20, no.~1, pp. 14--22, 2011.

\bibitem{morgan1990continuous}
N.~Morgan and H.~Bourlard, ``Continuous speech recognition using multilayer
  perceptrons with hidden markov models,'' in \emph{International conference on
  acoustics, speech, and signal processing}.\hskip 1em plus 0.5em minus
  0.4em\relax IEEE, 1990, pp. 413--416.

\bibitem{panayotov2015librispeech}
V.~Panayotov, G.~Chen, D.~Povey, and S.~Khudanpur, ``Librispeech: an asr corpus
  based on public domain audio books,'' in \emph{Proc. ICASSP}.\hskip 1em plus
  0.5em minus 0.4em\relax IEEE, 2015, pp. 5206--5210.

\bibitem{radford2022robust}
A.~Radford, J.~W. Kim, T.~Xu, G.~Brockman, C.~McLeavey, and I.~Sutskever,
  ``Robust speech recognition via large-scale weak supervision,'' \emph{arXiv
  preprint arXiv:2212.04356}, 2022.

\bibitem{zhang21ja_interspeech}
Y.~Zhang, E.~Bakhturina, and B.~Ginsburg, ``{NeMo (Inverse) Text Normalization:
  From Development to Production},'' in \emph{Proc. Interspeech 2021}, 2021,
  pp. 4857--4859.

\end{thebibliography}

\end{document}